\documentclass[reprint,amsmath,amssymb,aps,prb,showkeys]{revtex4-1}
\usepackage{graphicx}

\newcommand{\ket}[1]{\mathinner{|{#1}\rangle}}

\begin{document}

\title{Coherent control of the dynamics of single quantum-dot exciton qubit in a cavity}
\author{Antonio de Freitas}
\affiliation{Instituto de F\'isica, Universidade Federal de Uberl\^andia, 38400-902 Uberl\^andia, MG, Brazil}
\author{L. Sanz}
\affiliation{Instituto de F\'isica, Universidade Federal de Uberl\^andia, 38400-902 Uberl\^andia, MG, Brazil}
\author{Jos\'e M. Villas-B\^oas}
\email{villas-boas@infis.ufu.br}
\affiliation{Instituto de F\'isica, Universidade Federal de Uberl\^andia, 38400-902 Uberl\^andia, MG, Brazil}
\date{\today}

\begin{abstract}
   In this work we demonstrate theoretically how to use external laser field to control the population inversion of a single quantum dot exciton qubit in a nanocavity. We consider the Jaynes-Cummings model to describe the system, and the incoherent losses were take into account by using Lindblad operators. We have demonstrated how to prepare the initial state in a superposition of the exciton in the ground state and the cavity in a coherent state. The effects of exciton-cavity detuning, the laser-cavity detunings, the pulse area and losses over the qubit dynamics are analyzed. We also show how to use a continuous laser pumping in resonance with the cavity mode to sustain a coherent state inside the cavity, providing some protection to the qubit against cavity loss.
\end{abstract}

\pacs{78.67.Hc, 42.50.Ct, 42.70.Qs}
\keywords{quantum dot, cavity quantum eletrodynamics, solid state qubits, dynamical control}

\maketitle

\section{Introduction}
\label{sec:intro}
Quantum information processing (QIP) has become one of the most promising applications of quantum mechanics~\cite{Svore16,Nielsenbook}. The first criterion to be fulfilled for actual implementation of QIP~\cite{Divincenzo95,Bennett00} is the successful manipulation of a qubit, the basic unit for encoding quantum information~\cite{Schumacher95}. In this context, manipulation means coherent control of the quantum dynamics of the chosen qubit. The density matrix formalism is the perfect tool~\cite{Fano57} to explore those quantum dynamics, particularly when dealing with multipartite systems~\cite{Cohenbook}. For open quantum systems, being in contact with a reservoir, this formalism provides a theoretical environment to describe the action of decoherence~\cite{Schlosshauer05,Zurek03,Alivisatos96}.

Since the general state of one qubit is defined as a quantum superposition on a two-dimensional basis, it is natural to describe it as a $1/2$ spin system~\cite{Nielsenbook}. Coupling such a system with electromagnetic radiation is one of the paths to implement a coherent manipulation of the qubit, and the theoretical models to describe this interaction must take into the account the classical or quantum nature of the radiation~\cite{Scullybook}. When considering quantum radiation, the Jaynes-Cummings model~\cite{Jaynes63} is one of the most successful theoretical descriptions of the spin-boson interaction and subsequent dynamics~\cite{Shore93,Larson07,JCMJPBissue}.

Thinking about physical implementations, semiconductor nanostructures have become potential candidates for applications in QIP and quantum computing~\cite{Kane98,Hollenberg04}. Specifically, relevant contributions arise from the study of physical properties of quantum dots (QDs), often recognized as artificial atoms, in front of the discrete character of the energy spectrum due to the confinement of carriers~\cite{Burkard99,Loss98}. A qubit can be encoded inside a QD by using the charge~\cite{Shinkai09,Hayashi03}, the spin~\cite{Gao15,Press08,Petta05,Imamoglu99}, as well as excitonic states~\cite{Chen01,Xiaoqin03,Rolon10,Villas04} of the confined particle. One of the advantages of the QDs is the versatility on the experimental manipulation of (valence-conduction) band gap, with the subsequent customization of its optical properties.

The unique characteristics of quantum states in QD are behind the advantages of coupling a QD qubit with quantum light~\cite{Obrien09}, particularly involving nanocavities~\cite{Vahala03}. Once the QDs are created using semiconductor materials, it is possible to confine carriers in order to maximize the dipole-dipole interaction, coupling the nanostructure strongly with a chosen mode of electromagnetic field of the cavity~\cite{Yoshihiro03,Hennessy07}. The crescent interest on such an experimental setup lies on its potential for the miniaturization of the cavity quantum eletrodynamics (CQED), as found in atomic physics context~\cite{Haroche13}. Once a typical setup is smaller than a micrometer~\cite{Gao15,Obrien09}, one can think on the implementation of an \textit{on-chip} CQED~\cite{Lodahl15} using this kind of arrangement.

The rich optical response of quantum dots inside nanocavity includes nonlinearities~\cite{Bakker15} and non-trivial emission spectra~\cite{Majumdar12,Laucht09}. Several groups concentrate efforts on developing applications of QD-cavity setup as quantum light emitters\cite{Muller15,Heinze15,Munoz14,Farrow08,Kiraz04}, exploring quantum dynamics in a similar way of successful procedures for production of quantum states of light in atomic CQED~\cite{Deleglise08}. Other approach, directly related with QIP, is the use of quantum light for the coherent manipulation of the QD qubit. Typical phenomena of coherent dynamics as Rabi oscillations~\cite{Gao15,Dory16,Reithmaier04}, entanglement between a spin QD-qubit and photons~\cite{deGreve12,Schaibley13,Gao13} and exciton-photon entanglement~\cite{Muller14} have been observed on QD-cavity systems.

In this work, we propose encoding a qubit using an excitonic state, interacting with a  coherent state of light prepared in a nanocavity. The Jaynes-Cummings model is used to study the interaction between the quantum dot and the nanocavity and we consider photons losses in our treatment. We calculate the density matrix operator dynamics, once we are dealing with a multipartite open quantum system. Instead of a resonant condition between quantum dot and the cavity~\cite{Dory16}, we assume a non-resonant, self-trapped dynamics.

To create and maintain a coherent state inside the cavity we use a continuous laser field applied in resonance with the cavity mode. To control the qubit rotations, the coherent manipulation of the system is done by applying external laser pulses. To find the best set of parameters for the coherent manipulation, we calculate the average occupation of the exciton state as a function of two detunings: laser-cavity and exciton-cavity. The effect of the pulse area is also evaluated and the residual effect of the pulses over the cavity state is surveyed by checking the average number of photons and photon distribution. We explore the behavior of the QD qubit by studying the population inversion, the dynamics over the Poincar\'{e} sphere and the purity of qubit, the last one using Von Neumann entropy.

\section{Theory}
\label{sec:theory}
Our system is composed of a QD, treated here as a two-level system, coupled to a single mode nanocavity. To explore this physical setup, we use the Jaynes-Cummings model~\cite{Jaynes63} under the Rotating Wave Approximation (RWA)~\cite{Knight73,Brown00,Irish07,Laucht09}. To model the external lasers we use the dipole and RWA approximations \cite{Scullybook} and assume that the pulsed laser only interact with the QD while the continuous laser only with the cavity. The Hamiltonian can be written as ($\hbar=1$)
\begin{align}
H=&\omega_x\sigma_+\sigma_-+\omega_c a^{\dagger}a + g(\sigma_+ a + \sigma_- a^{\dagger})\nonumber\\
&+\frac{\Omega(t)}{2}(e^{-i\omega_p t}\sigma_+ + e^{i\omega_p t}\sigma_-)\nonumber\\
&+J(e^{-i\omega_l t}a^{\dagger}+e^{i\omega_l t}a),
\label{eq:hamiltonian}
\end{align}
where $\sigma_{\pm}$ are the pseudospin operators for the QD exciton qubit, $a^{\dagger}$ and $a$ are the creation and annihilation operators for photons inside the cavity, $\omega_x$, $\omega_c$, $\omega_p$ and $\omega_l$ are the frequencies of exciton, cavity mode, pulsed laser and continuous laser field, respectively, $\Omega(t)=\mu E(t)/\hbar$ is the Rabi frequency that describes the
exciton-laser interaction, with $\mu$ being the electric dipole strength of the exciton transition and $E(t)$ the amplitude of the electric field of the laser, which can be constant or have different shapes in the pulsed excitation, and $J$ contains information about the laser field amplitude and the cavity transmission coefficient.

We encode a quantum bit using the QD exciton state, being $\ket{0}$ the state with no exciton and $\ket{1}$ the exciton state, and the cavity is described by the usual Fock basis, $\ket{n}$. The Hamiltonian basis is depicted as $\ket{i,n}$ with $i=0$ or $1$, indicating the state of the QD qubit and $n$ being the number of photons in the cavity.

To obtain the dynamics of our physical system, we numerically solved the time dependent density matrix in the Lindblad form
\begin{align}
\frac{d\rho}{dt}&=-\imath[H,\rho]+\kappa\mathcal{D}[a]+\gamma\mathcal{D}[\sigma_-]+\phi\mathcal{D}[\sigma_z]
\end{align}
where $H$ is the full Hamiltonian [Eq.\  \eqref{eq:hamiltonian}] and $\mathcal{D}[L]=L\rho L^\dagger-\frac{1}{2}(L^\dagger L \rho + \rho L^\dagger L )$ is the Lindblad superoperator, which contains the incoherent terms of the density matrix and assumes a Markovian approximation. Here $\kappa$ is the photon loss rate of the cavity, $\gamma$ is the decay rate of the QD and $\phi$ is the pure dephasing rate of the QD.

In order to get a deeper understanding on the QD qubit dynamics we use the Poincar\'{e}-sphere representation, an analogous of the Bloch-sphere representation for mixed states \cite{Dory16}. The use of this representation is common in the context of cavity quantum electrodynamics~\cite{Rempe15}. The QD qubit treated here is not in a pure state: when losses are not considered ($\kappa$ and $\gamma$ null), the QD qubit is one of the parts of a bipartite system, and has some degree of entanglement with the cavity. For the contrary, when considering losses, one can say the QD qubit exchanges just a portion of its information with the cavity, once the bipartite system is now open. To obtain the components $X$, $Y$ and $Z$ of the Bloch vector we first compute the reduced density operator for the QD qubit by doing the partial trace over the cavity variables
\begin{align}
\hat{\rho}^{\mathrm{QD}}=\mathrm{Tr}\left[\hat{\rho}\right]_{\mathrm{cav}}.
\end{align}
then
\begin{align}
X&=2\mathrm{Re}(\hat{\rho}^{\mathrm{QD}}_{01})\nonumber\\
Y&=2\mathrm{Im}(\hat{\rho}^{\mathrm{QD}}_{10})\nonumber\\
Z&=\hat{\rho}^{\mathrm{QD}}_{00}-\hat{\rho}^{\mathrm{QD}}_{11}.
\label{eq:blochvec}
\end{align}
On Poincar\'{e} sphere, the $Z$ component of the mixed, Block-like, vector coincides with the population inversion. The value of azimuthal angle is known as relative phase which corresponds to $\phi$ in the pure qubit represented as $C_0\ket{0}+e^{i\phi}C_1\ket{1}$, with $C_i$ being related to the population of the state $\ket{i}$. A Bloch vector with a null azimuthal angle lies on the $YZ$ plane.\cite{Nielsenbook}

To quantify the purity of the QD qubit, we use the Von Neumann entropy $S$, which is defined as  \cite{Nielsenbook}
\begin{align}
S(\hat{\rho}^{\mathrm{QD}})=-\mathrm{Tr}[\hat{\rho}^{\mathrm{QD}}\log_2(\hat{\rho}^{\mathrm{QD}})].
\label{eq:entropy}
\end{align}
From the definition, $S=0$ indicates that the QD qubit is a pure system, described by a state separated from the cavity and the reservoir. If losses are not considered on the description of the system, maximal degree of entanglement between the qubit with the cavity corresponds to $S=1$. 

\section{Results and discussion}
\label{sec:results}

To analyze the dynamics of QD qubit we parameterize all frequencies in units of $g$, so it is easy to convert the values obtained here to real experimental situations. For example, in photonic crystal nanocavity $\hbar g$ is of the order of $0.1$ meV \cite{Laucht09}. This parametrization also allows our results to be applied to other cavity systems, which we will not discuss here. To solve the time-dependent master equation we carefully choose the size of photon basis to describe accurately the cavity state according to its mean occupation, taking into account the effects of interaction between QD, cavity and external lasers. In most of the cases studied, a Fock basis of $n_\mathrm{max}=70$ was sufficient, so we set this value for all simulations presented here.

To get some insight to what we can do, lets first ignore incoherent effect ($\kappa,\gamma,\phi=0$) and consider the initial state of our system as a direct product of a coherent state $\ket{\alpha}$ in the cavity and QD in the ground state $\ket{0}$, thus, $\ket{\Psi(0)}=\ket{0}\ket{\alpha}$. The coherent state can be represented in the basis of Fock states as
\begin{align}
\ket{\alpha}=\exp(-|\alpha|^2/2)\sum_n\frac{\alpha^n}{\sqrt{n!}}\ket{n},
\label{eq:pn}
\end{align}
where $|\alpha|^2=\langle n\rangle$ is mean number of photons. A coherent state in the cavity can be created by sending a laser in resonance with the cavity mode for a short period of time ($J\neq 0$ and $\omega_j=\omega_c$ in our Hamiltonian).

\begin{figure}[!h]
\centering\includegraphics[width=1.0\linewidth]{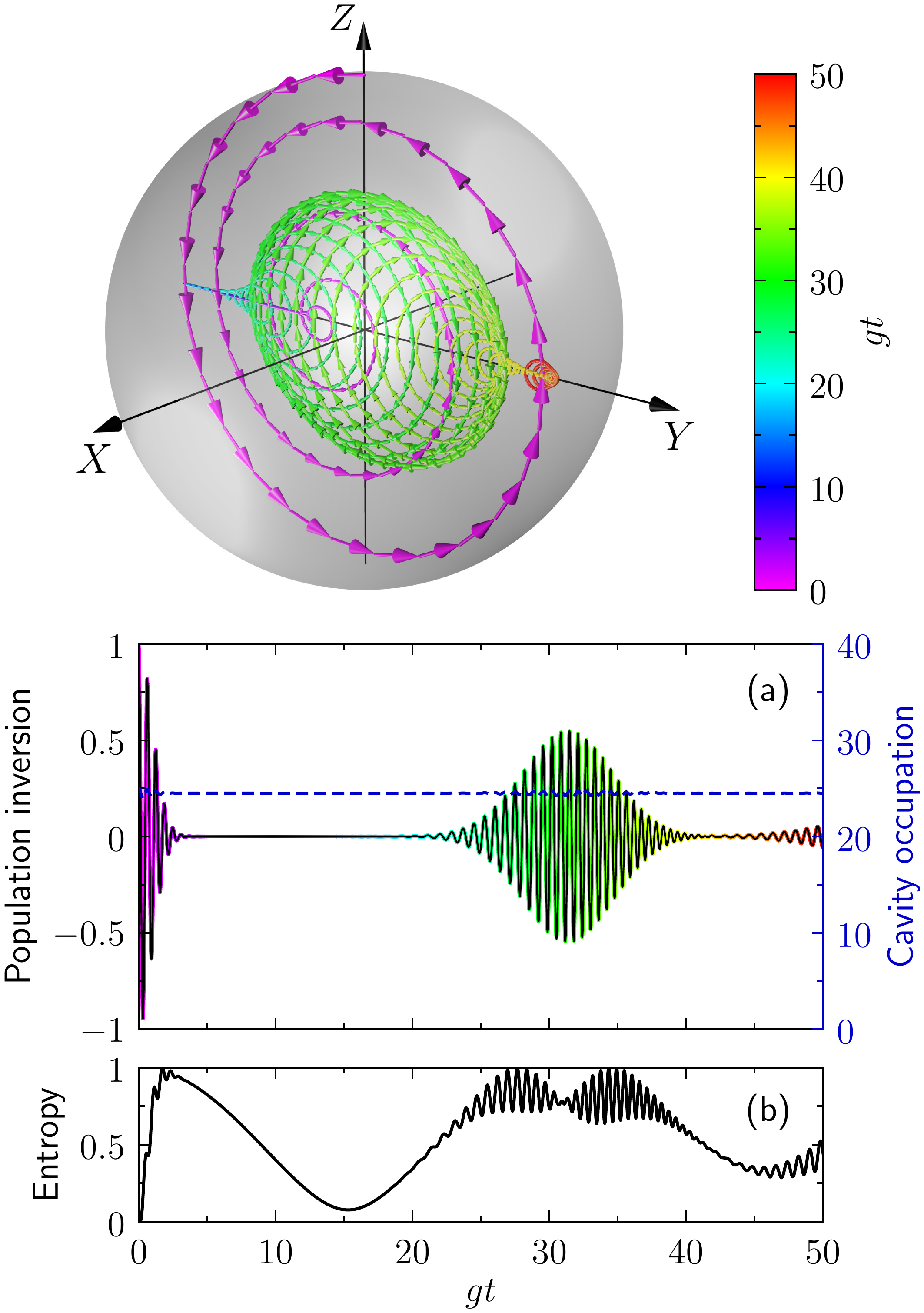}
\caption{(color online) Dynamics of a QD qubit inside a cavity considering full resonant condition given by $\delta_x=0$ and in the absence of lasers fields. (a) Population inversion and cavity occupation (dashed blue line) as a function of $gt$. (b) Von Neumann entropy as a function of $gt$. Upper panel shows the evolution of QD qubit over the Poincar\'e sphere.}
\label{fig:resonant}
\end{figure}

Figure~\ref{fig:resonant} illustrates several aspects of the QD qubit dynamics when interacting with the cavity without the action of any external laser field ($J,\Omega=0$) and neglecting incoherent effect ($\kappa,\gamma,\phi=0$). We consider that the exciton state is resonant with the cavity mode, $\delta_x=\omega_x-\omega_c=0$, and an initial cavity state with $\langle n\rangle=25$. Figure~\ref{fig:resonant}(a) shows the behavior of the population inversion, $Z(t)$, with the apparition of collapses and revivals around an average value $\langle Z\rangle=0$, in agreement with the predictions of the Jaynes-Cummings model. 
Notice that the average occupation of the cavity stays almost constant at a value around $\langle n\rangle\simeq 25$.

The dynamics of the QD qubit on Poincar\'{e} sphere is shown in the upper panel of Fig.\ \ref{fig:resonant}, illustrated by constructing a vector which components are defined by Eqs.\ \eqref{eq:blochvec}. In this figure, a scale of colors and arrows helps to visualize the temporal evolution. First, the QD qubit starts from the pure state $\ket{0}$, indicated by a purple curves at the north pole of the Poincar\'{e} sphere. Then, at short times, the QD qubit performs a rotation around the $X$ axis, which is restricted to the $YZ$ plane ($X=0$), showing that the relative phase is null under the evolution. It is also observed that the norm shrinks as time increases, which becomes evident at long times, as shown by green/yellow/red arrows.

The QD qubit dynamics restricted to a small region inside the Poincar\'{e} sphere, is connected with a high degree of entanglement with the cavity. This behavior is better understood by checking the evolution of the Von Neumann entropy $S(\hat{\rho}_{\mathrm{QD}})$, Eq.\ \eqref{eq:entropy}, shown in Fig.\ \ref{fig:resonant}(b). The initially pure QD qubit ($S=0$ at $gt=0$) performs oscillations between high and low entangled states, with an almost complete purification at $gt\simeq15$ associated with the collapse on population inversion. This behavior repeats subsequent collapses (not shown). As time further increases, the degree of entanglement increases and approaches the maximal entangled states indicated by $\mathcal S(\hat{\rho}_{\mathrm{QD}})\simeq 1$.

\begin{figure}[!h]
\centering\includegraphics[width=1.0\linewidth]{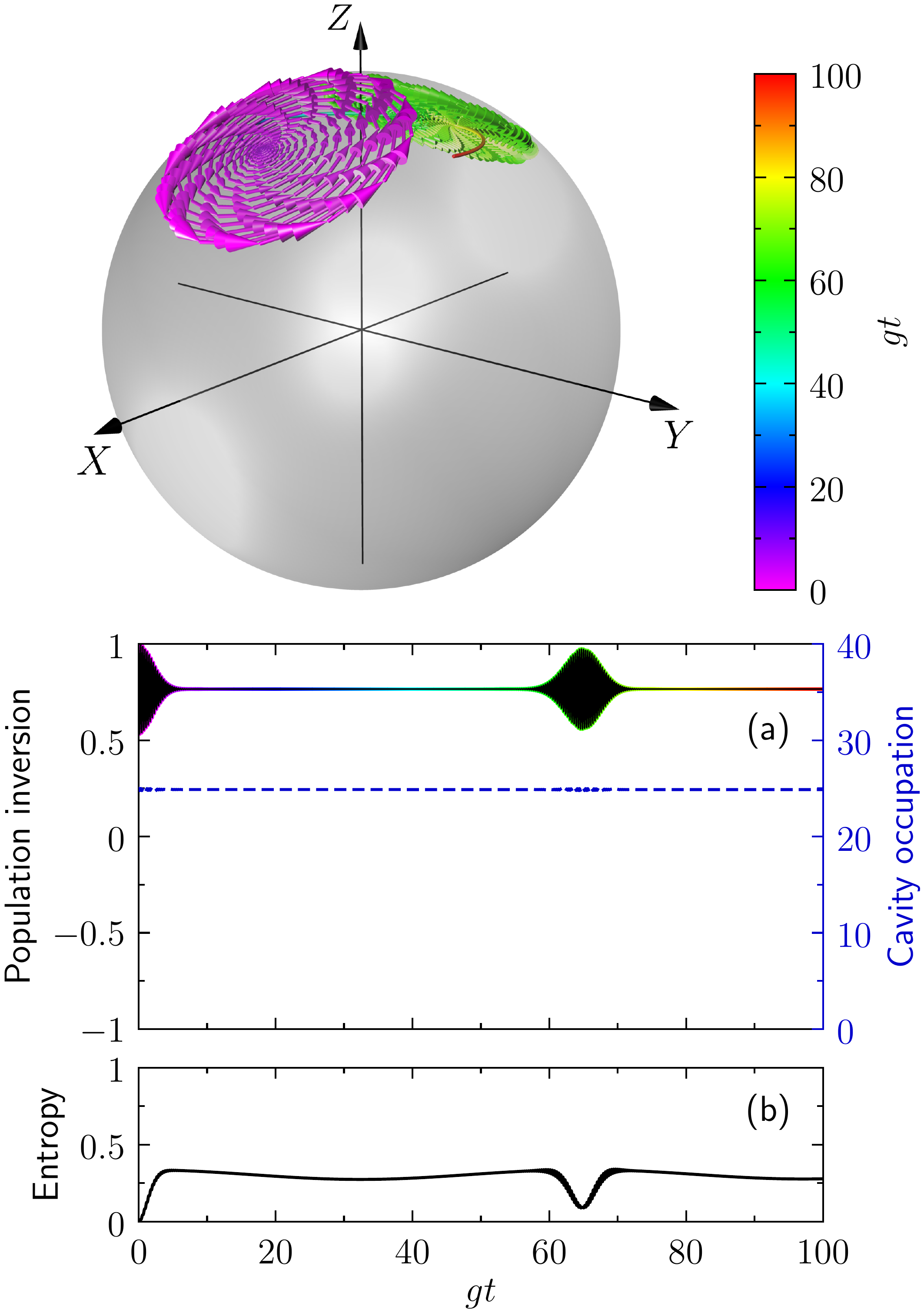}
\caption{(color online) Dynamics of a QD qubit inside a cavity considering non-resonant condition with $\delta_x=-18g$ and in the absence of lasers fields. (a) Population inversion and cavity occupation (dashed blue) as a function of $gt$. (b) Von Neumann entropy as a function of $gt$. Upper panel shows the evolution of QD qubit over the Poincar\'e sphere.}
\label{fig:nonresonant}
\end{figure}
It is also important to explore the behavior of the system under a non-resonant condition between QD qubit and cavity. Figure~\ref{fig:nonresonant} shows the same theoretical tools used on the description of the QD qubit resonant dynamics but considering $\delta_x=-18g$. Figure~\ref{fig:nonresonant}(a) shows that the population inversion does not perform complete oscillations between the QD qubit states, although collapses and revivals around the average value $\langle Z\rangle \simeq0.75$ are still present. This phenomena is called \textit{self-trapping} or population trapping and is a well-know aspect of non-resonant dynamics concerning two-level systems. The average occupation of the cavity stay constant at the value of $\langle n\rangle=25$ over the evolution as seen by the blue line linked to the right axis. The self-trapping is more evident in the upper panel of Fig.\ \ref{fig:nonresonant}, once the mixed state vector is confined in a restricted region on the north hemisphere of Poincar\'{e} sphere. It is worth noting that the mixed vector performs rotations around $Z$ axis. Thinking in terms of a general qubit state written as $\ket{\Psi}= C_0\ket{0}+e^{i\phi}C_1\ket{1}$, changes in the relative phase $\phi$ are connected with changes with the value of azimuthal angle of the qubit Bloch vector. Thus, the non-resonant dynamics brings a gain of relative phase of the mixed vector, which oscillated between $0$ and $2\pi$. Figure \ref{fig:nonresonant}(b) shows the Von Neumann entropy, where differently from Fig.~\ref{fig:resonant} the entanglement of the QD qubit suffers a stabilization, having asymptotic value around $S\simeq 0.3$. Notice also that an increase of the purification appears at the revival of oscillations in the population inversion, indicating that the system try to recover its initial state. In general, nonresonant condition preserves the purity of QD qubit, once it prevents the qubit to interact in an efficient way with the cavity.

\begin{figure}[!h]
\centering\includegraphics[width=1.0\linewidth]{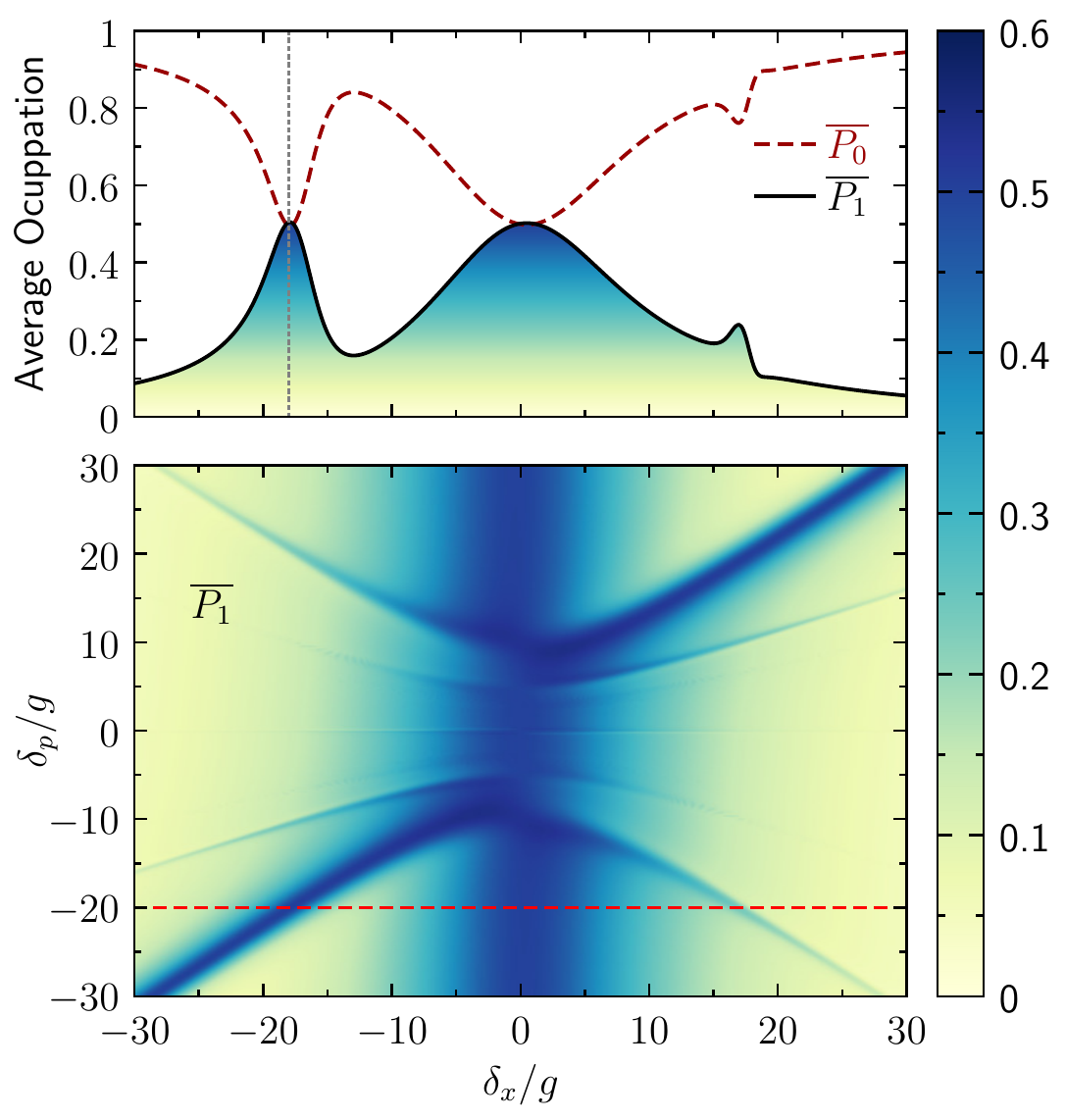}
\caption{(color online). Lower panel: False color plot of the average occupation of the exciton states $\overline{P_{1}}$ after the system being prepared in the state $\ket{\Psi(0)}=\ket{0}\ket{\alpha}$ for a constant laser with $\Omega=2g$ as functions of detunings $\delta_x=\omega_x-\omega_c$ and $\delta_p=\omega_p-\omega_c$ with $J=0$. Dark regions correspond to energy configurations where the qubit exciton state $\ket{1}$ is being populated. Dashed red line illustrates the condition $\delta_p=-20g$, where individual average occupations are shown in the upper panel. Dotted gray line in the upper panel correspond to $\delta_x=-18g$, being the best condition for population inversion.}
\label{fig:averpop}
\end{figure}
After checking the basic aspects on QD qubit-cavity dynamics, we are ready to understand the action of the external lasers and incoherent effects. The main problem here is to find the right parameters to send the laser pulses as the coupling between QD and cavity modifies its interaction with the laser. A practical and fast way for survey the effect of this new ingredient can be done through calculations of the average occupation $\overline{P_{i,n}}=\lim_{\tau\to\infty} \frac{1}{\tau}\int_{0}^{\tau} P_{i,n}(t)dt$ of basis states $\ket{i,n}$ as a function of some physical parameters neglecting incoherent process and assuming a constant laser excitation ($\Omega(t)$ constant), which allow us to use the time independent Schr\"{o}dinger equation to compute the time evolution. As we are interested in the qubit dynamics and wants to investigate where the population inversion occurs, we write the average occupation of the $i^{th}$ QD qubit state as $\overline{P_{i}}=\sum_{n}\overline{P_{i,n}}$. Here we choose to seek for best exciton-cavity $\delta_x=\omega_x-\omega_c$, and laser-cavity $\delta_p=\omega_p-\omega_c$ detunings and the reason for this choice is that the cavity frequency $\omega_c$ is usually fixed by construction, making our procedure very similar to what one would do in an experimental setup.

Figure~\ref{fig:averpop} (lower panel) shows the average occupation of the exciton state as functions of $\delta_x$ and $\delta_p$ for our system under a constant laser excitation with $\Omega=2g$ interacting with the QD, keeping $J,\kappa,\gamma,\phi=0$. The initial state is again $\ket{\Psi(0)}=\ket{0}\ket{\alpha}$ with $\langle n\rangle=25$. Bright regions indicate low population of the exciton state in opposition to dark areas, therefore, dark areas are good candidates for laser induced population inversion. In the upper panel of Fig.\ \ref{fig:averpop} we show a cut in the false color plot for $\delta_p=-20g$ (red dashed line in the lower panel), showing the average occupation of the two QD qubit components. Gray dotted lines at $\delta_x=-18g$ indicates the maximum value of $\overline{P_{1}}$, being a good candidate to present a high degree of population inversion under pulsed excitation.

\begin{figure}[!h]
\centering\includegraphics[width=1.0\linewidth]{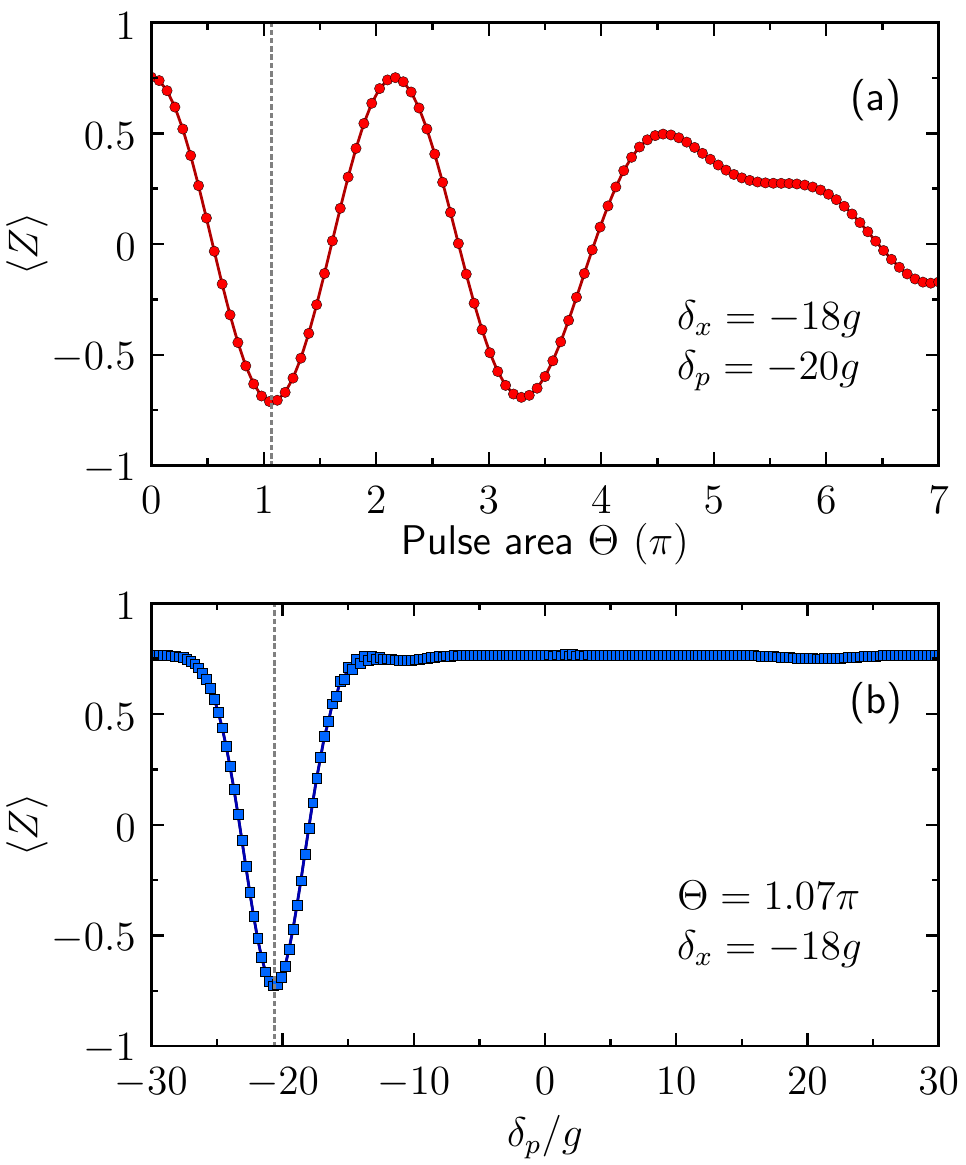}
\caption{(color online) (a) Average of the population inversion $\langle Z\rangle$ as a function of the pulse area $\Theta$ after the application of a single Gaussian pulse at $gt_1=10$ with duration of $gt_p=0.7$ and detunings $\delta_x=-18g$ and $\delta_p=-20g$ . (b) Average of the population inversion $\langle Z\rangle$ as a function of $\delta_p$ for a pulse with same parameters as in (a) with an area $\Theta=1.07\pi$.}
\label{fig:Central_av}
\end{figure}

In order to gain a fine control of the dynamics in realistic situations, we check the effect of the pulse parameters more closely. We proceed the simulations using a single Gaussian pulse of duration of $gt_p=0.7$, which in typical strong-coupled QD-cavity system would be a pulse with duration of the order of 4 to 5 ps, centered at $gt_p=10$, with optimal parameters obtained from Fig.~\ref{fig:averpop} ($\delta_x=-18g$ and $\delta_p=-20g$). We observed that the average of the population inversion $\langle Z\rangle$ after the pulse is strongly dependent of the pulse area, defined as $\Theta=\int_{-\infty}^{\infty}\Omega(t)dt$, as we can see in Fig.\ \ref{fig:Central_av}(a). The maximum change in $\langle Z\rangle$ is obtained for $\Theta\simeq1.07\pi$ and $\Theta\simeq3.3\pi$. The dependence of $\langle Z\rangle$ with $\Theta$ shows that the self-trapped dynamics, initially fixed at some polar angle over Poincar\'e sphere, can be set up on demand by the action of a pre-arranged laser pulse. It is interesting to note that high values of pulse area bring nonlinear effects, linked with the increase of the pulse intensity and resulting in a lack of control of the dynamics. This is indicated by nontrivial values of $\langle Z\rangle$ for $\Theta>4$.

We now simulate the behavior of $\langle Z\rangle$ as a function of the laser detuning to the cavity frequency considering a laser pulse with area $\Theta=1.07\pi$, corresponding to the first minimum in Fig.\ \ref{fig:Central_av}(a). As we can see in Fig.\ \ref{fig:Central_av}(b), $\langle Z\rangle$ changes drastically for values around $\delta_p=-20g$, in accord with the predictions for continuous laser as shown in Fig.\ \ref{fig:averpop}. It is also interesting to notice that there is a dip of $\langle Z\rangle$ around $\delta_p=-20.6g$, in a range of $\approx 5\delta_p$, meaning the effects of the pulse on dynamics permits some flexibility over the exact value of $\delta_p$. For practical applications, once it is established an adequate value of $\delta_p$ (using the survey of populations), a specific pulse can be set up on demand in order to perform controlled self-trapped dynamics of the QD qubit. This set of parameter, $\delta_x=-18g$, $\delta_p=-20.6g$ and $\Theta\simeq1.07\pi$, will be kept constant for the rest of the paper.

To corroborate our approach, we proceed to explore the dynamics of population inversion, QD qubit on Poincar\'e sphere and Von Neumann entropy. We discuss the action of a sequences of three pulses. We consider again the QD qubit initialized at the ground state $\ket{0}$ and the cavity in the coherent state $\ket{\alpha}$ with $\langle n\rangle=25$ and the conditions described before.

\begin{figure}[!h]
\centering\includegraphics[width=1.0\linewidth]{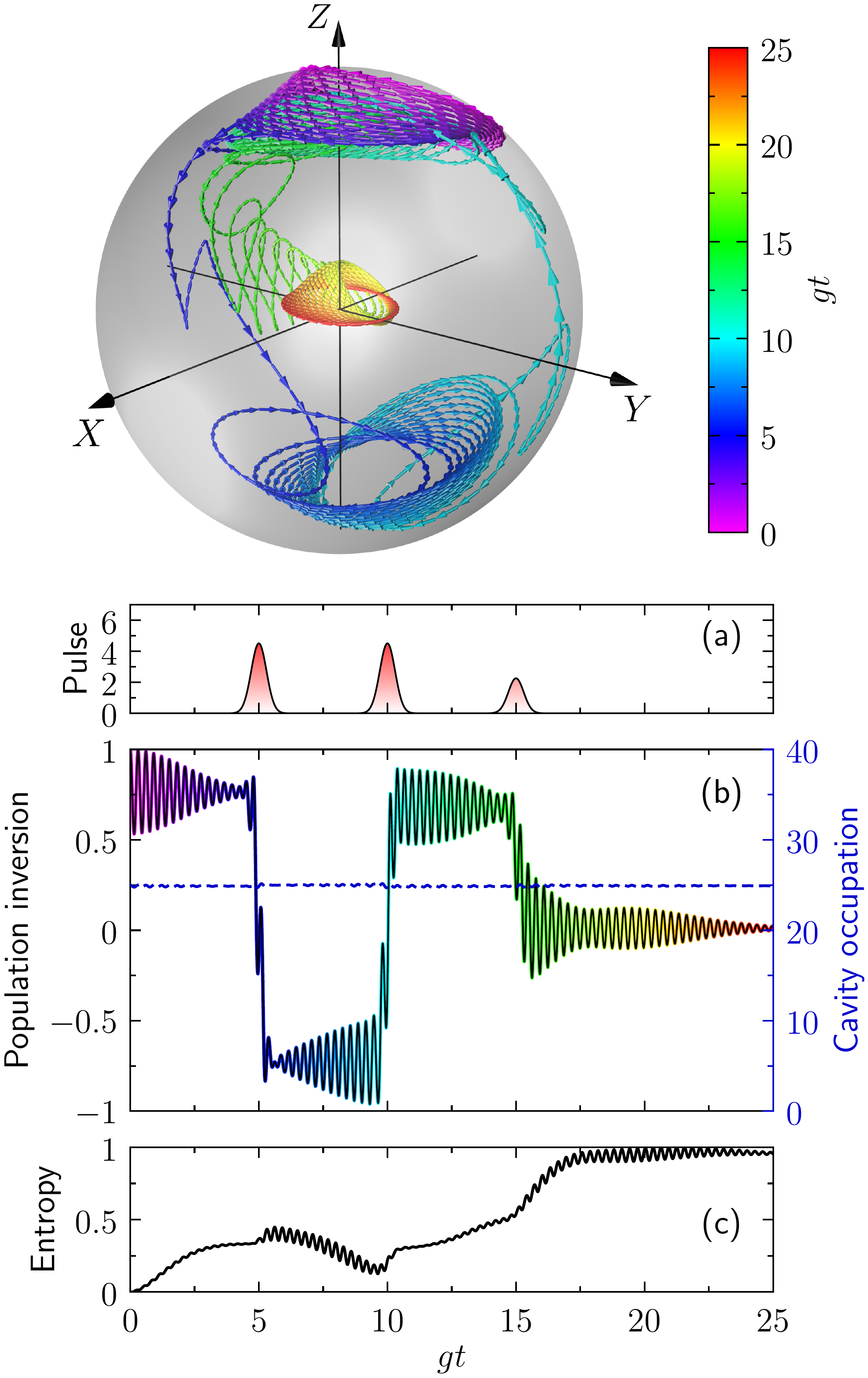}
\caption{(color online) Dynamics of a QD qubit inside a cavity considering $\delta_x=-18g$, $\delta_p=-20.6g$, neglecting losses and under the effect of three Gaussian, the first and second with the same pulse area $\Theta=1.07\pi$, and the last with $\Theta=1.07\pi/2$, applied at $gt_1=5$, $gt_2=10$ and $gt_3=15$, all with duration of $gt_p=0.7$. (a) Sequence of pulses used in the simulation, (b) Population inversion and cavity occupation (dashed blue) as a function of $gt$ and (c) Von Neumann entropy as a function of $gt$. Upper panel shows the evolution of QD qubit over the Poincar\'e sphere, with the color code and arrows indicating the time sequence.}
\label{inv2}
\end{figure}

Figure~\ref{inv2} shows the QD qubit dynamics under the action of a sequence of three pulse, all with duration of $gt_p=0.7$ for the best detuning found before. The format of the pulse is shown in Fig.\ \ref{inv2}(a). The population inversion, Fig.\ \ref{inv2}(b), shows that the QD qubit initially (for $gt<5$) performs self-trapped oscillations around $\langle Z\rangle\simeq0.75$. This can be better visualized in the top panel of Fig.\ \ref{inv2}, where the dynamics is restricted to the north hemisphere of the Poincar\'{e} sphere. After the application of the first pulse at $gt_1=5$ with pulse area $\Theta_1=1.07\pi$, the population inversion starts an oscillation with a new structure of collapses and revivals, oscillating around $\langle Z\rangle\simeq-0.73$, still in a self-trapped dynamics as can also be seen in the Poincar\'{e} sphere, moving the oscillation from north to the south hemisphere. After the application of the second pulse at $gt_2=10$ with same pulse area $\Theta_2=1.07\pi$, the self-trapping oscillations changes to an average value around $\langle Z\rangle\simeq0.72$, moving back to the north hemisphere of the Poincar\'{e} sphere. Further control can be obtained with the last pulse, applied at $gt_3=15$ with a pulse area half of the previous case ($\Theta_3=1.07\pi/2$), creating a superposition between exciton and ground state, showing that we can control the QD qubit state reasonable well with this choice of pulse sequence.
Concerning the entanglement dynamics, again explored using the Von Neumann entropy plotted in Fig.~\ref{inv2}(b), it is worthy noting that the two first pulses changes slightly the entanglement degree of QD qubit, keeping the value of the entropy below $0.5$. The third pulse, on the other hand, creates a situation similar to the full resonant case,  Fig.\ \ref{fig:resonant}, where the entropy goes to maximum, and the inversion exhibits oscillations around the average value of $\langle Z\rangle=0$.

One advantage of using non-resonant condition is that the QD lifetime increases from a few hundred of picoseconds to a few nanoseconds due to the reduction of fluctuations in the vacuum of the electromagnetic field \cite{Hennessy07}. Taking the lower limit, lets assume that lifetime of the QD is about $0.7$ ns, which gives a decay rate of the order of $\gamma=0.01g$. Since we are proposing the manipulation the QD qubit in a timescale of a few picoseconds, the effects of this decay rate can be neglected from our analysis. We also consider a pure dephasing rate of the same order of magnitude $\phi=0.01$ and as we shall see later, pure dephasing plays an important role in the preparation of the initial stated used in dynamics of the coupled system.

\begin{figure}[!h]
\centering\includegraphics[width=1.0\linewidth]{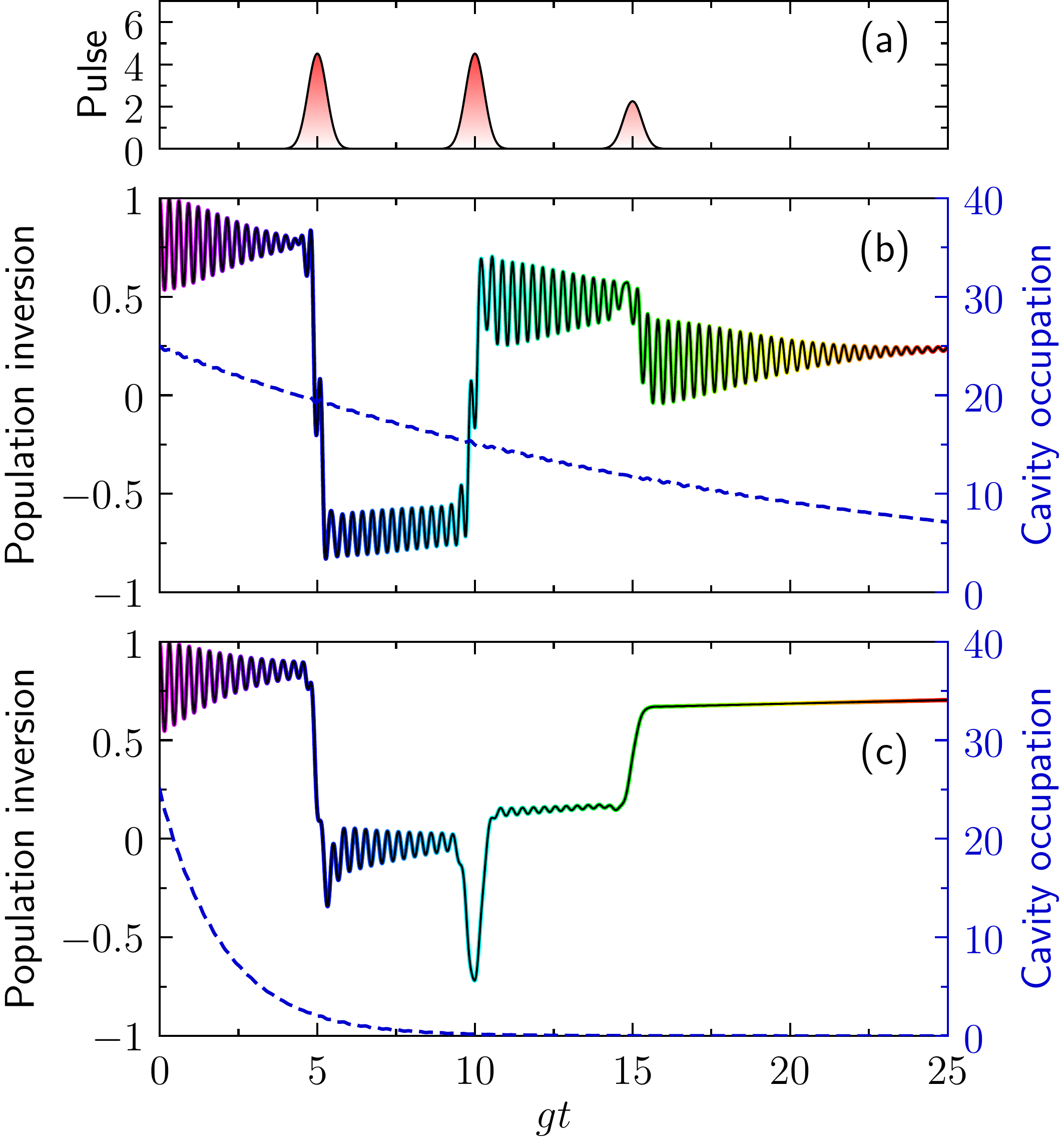}
\caption{(color online) Dynamics of a nonresonant QD qubit inside a cavity under the effect of three Gaussian pulses for the same set of parameters used in Fig.\ \ref{inv2}, but now considering decoherence. (a) Sequence of pulses used in the simulation. (b) Population inversion and cavity occupation (dashed blue line) as a function of $gt$ for $\kappa=0.05g$, $\gamma=0.01g$ and $\phi=0.01g$. (c) Same as (b) for $\kappa=0.5g$.}
\label{inv2ab}
\end{figure}

Cavity loss is another term that can not be neglected since the average life span of a photon within a cavity may be as small as a few picoseconds in a bad cavity. To understand the effects of the cavity loss, in Fig.~\ref{inv2ab} we show the dynamics of the system using same parameters and sequence of pulses as in Fig.~\ref{inv2}, but now including incoherent effects. In Fig.~\ref{inv2ab}(b) we use $\gamma=0.01g$, $\phi=0.01g$ and $\kappa=0.05g$, while in Fig.\ \ref{inv2ab}(c) we use $\gamma=0.01g$, $\phi=0.01g$ and $\kappa=0.5g$. As we can see in this figure, even a small loss of the cavity is enough to break the QD qubit manipulation. Notice that the average occupation of the cavity photons decease exponentially and we have $\langle n\rangle\simeq20$, $\langle n\rangle\simeq15$ and $\langle n\rangle\simeq12$ when the first, second and third pulses are applied, respectively. Pure dephasing and exciton spontaneous decay plays no role in this particular case as the time scale is small. Notice that we have used $(\ket{\Psi(0)}=\ket{\alpha}\ket{0})$ as initial state, which is a pure state. For the case of $\kappa=0.5g$, Fig.\ \ref{inv2ab}(c), the occupation of the cavity goes to zero very quickly, being zero before the second pulse, and the coherent manipulation of the QD qubit is completely destroyed. Thus, even thought the cavity is out of resonance, the loss of the cavity plays an important role on the QD qubit manipulation.

It is important to mention that the coherent state under cavity loss will still be a coherent state, but with a lower average of photons. The main problem here is that the resonance condition (best parameter we found in Figs. \ref{fig:averpop} and \ref{fig:Central_av}) changes over time as the photon population decreases and a complete control of the QD state will require an previous evaluation of the best parameter for each pulse of the pulse sequence to control its states. Experimentally, this is challenging, as it require sending pulses with different frequency and intensities in a short time scale.

\begin{figure}[!h]
\centering\includegraphics[width=1.0\linewidth]{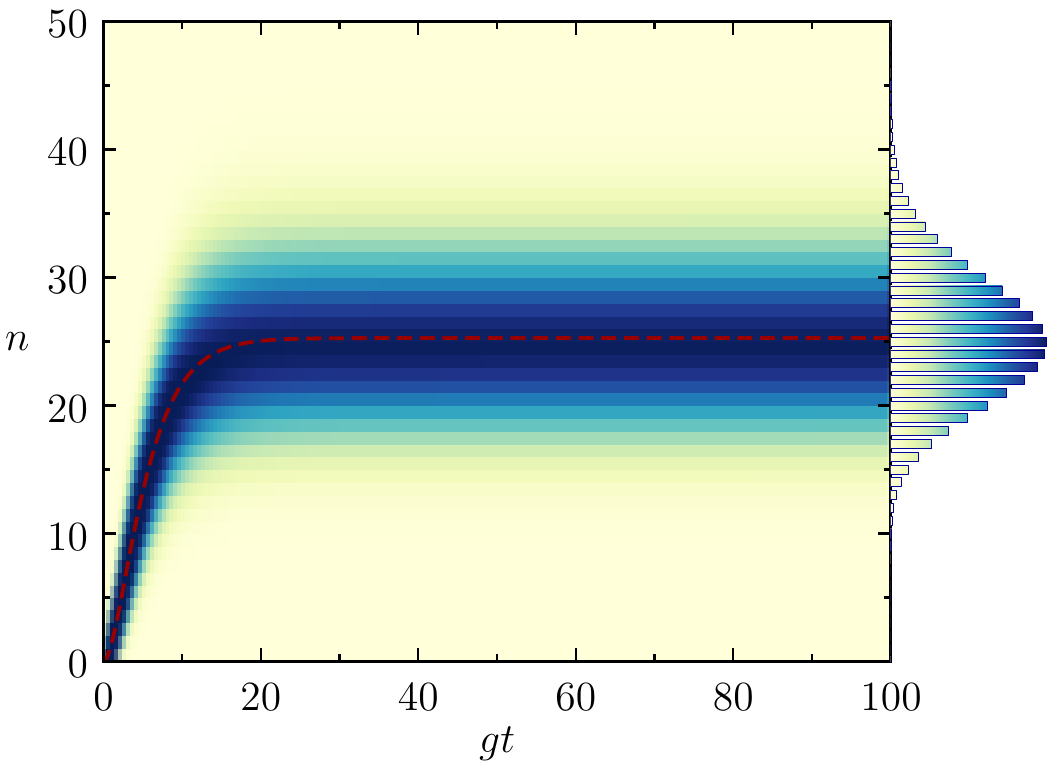}
\caption{(color online) (a) Evolution of the photon distribution in the cavity for $\kappa=0.5g$, $\gamma=0.01g$, $\phi=0.01g$ and $J=1.28g$ assuming $\ket{\Psi(0)}=\ket{0}\ket{0}$ as initial state. Right panel shows the photon distribution for the quantum state at $gt=100$.}
\label{coh1}
\end{figure}

To solve this problem, instead of using the laser described by the coupling $J$ to only prepare the initial coherent state in the cavity, lets keep it as a constant pump to maintain a steady coherent state in the cavity. In Fig.~\ref{coh1} we plot the evolution of the photon distribution in the Fock basis considering $\delta_x=-18g$, $\delta_p=-20.6g$, (same parameters used before), $\omega_j=\omega_c$, $J=1.28g$, $\kappa=0.5g$, $\gamma=0.01g$ and $\phi=0.01$ as a function of $gt$, assuming $\ket{\Psi(0)}=\ket{0}\ket{0}$ as initial state. Here we choose $j=1.28g$ because that produces an steady coherent state with $\langle n\rangle\simeq 25.3$, close to the condition used in Fig.\ \ref{inv2}. The right panel is a view of the photon distribution for $gt=100$, showing a coherent photon distribution.

\begin{figure}[!h]
\centering\includegraphics[width=1.0\linewidth]{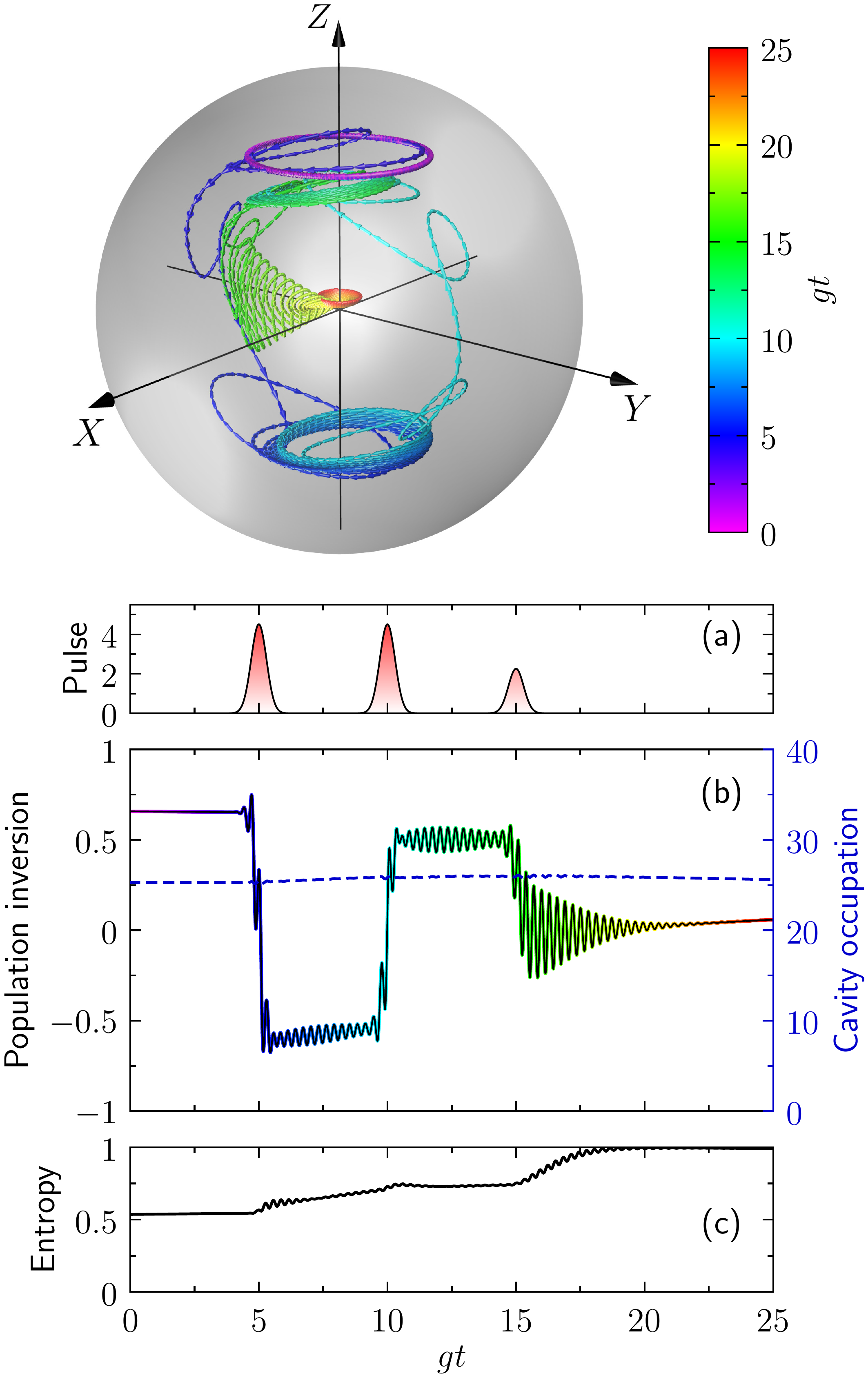}
\caption{(color online) Dynamics of a QD qubit inside a cavity under the effect of three Gaussian pulses as in Fig.\ \ref{inv2}, now considering $\kappa=0.5g$, $\gamma=0.01g$, $\phi=0.01g$ and $J=1.28g$ and the initial state the state at $gt=100$, whose photon distribution is shown in the right panel of Fig.\ \ref{coh1}.  (a) Sequence of pulses used in the simulation, (b) Population inversion and cavity occupation (dashed blue line) as a function of $gt$. (c) Von Neumann entropy as a function of $gt$. Upper panel shows the evolution of QD qubit over the Poincar\'e sphere, with the color-code indicating the time sequence.}
\label{inv2d}
\end{figure}

Using the state at $gt=100$ obtained with $J=1.28g$ as initial state of the system, in Fig.\ \ref{inv2d} we shows the dynamics of the system under the same pulse sequence and parameters as in Fig.\ \ref{inv2ab}(c), with additional parameters $\omega_j=\omega_c=0$, $\kappa=0.5g$, $\gamma=0.01g$ and $\phi=0.01g$. As we can see in Fig.\ \ref{inv2d}(b), we have now a better control the inversion of population with our pulse sequence as the cavity occupation is kept almost constant over the evolution despite of the application of the lasers pulses interacting with the QD. The value is very close to our ideal situation (with no losses), with an average of photons in the cavity around $\langle n\rangle\simeq 25$ as we initially planed. Notice, however that the range of population inversion is decreased, this is due to the pure dephasing rate acting in the preparation of the state of cavity (the evolution during the pulse sequence is too short for the pure dephasing and decay of the exciton to play a role). The effect of the pure dephasing rate in the initial state can also be seen is in Fig.\ \ref{inv2d}(c), where the initial value of the Von Neumann entropy is about $0.6$. Despite the initial difference, the evolution of the Von Neumann entropy is very similar to the case analyzed previously in Fig.\ \ref{inv2}(c).

\begin{figure}[!h]
\centering\includegraphics[width=1.0\linewidth]{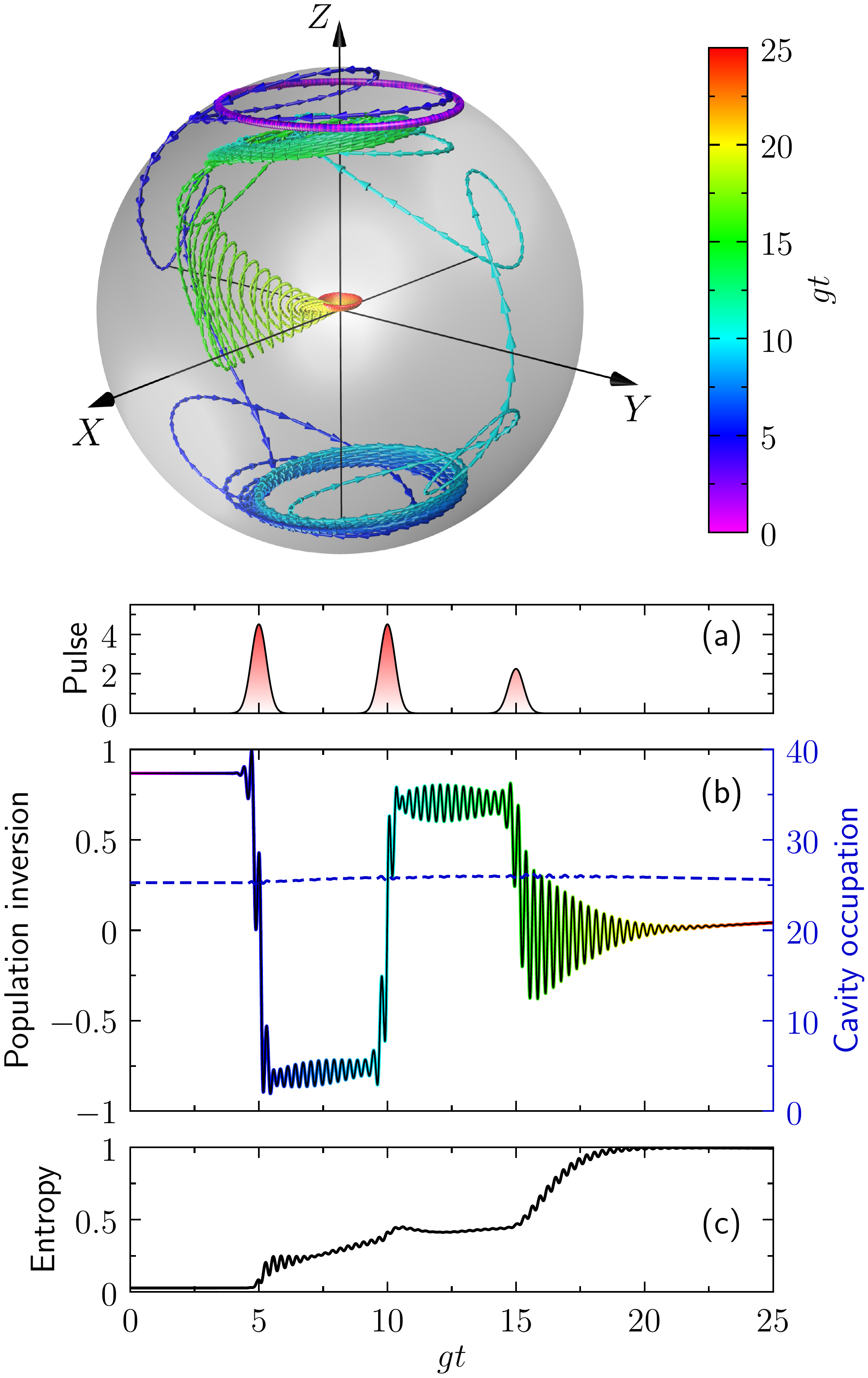}
\caption{(color online) Dynamics of a QD qubit inside a cavity under the effect of three Gaussian pulses and parameters as in Fig.\ \ref{inv2d}, now neglecting pure dephasing and considering the initial state as the state at $gt=100$ for $\phi=0.0$.  (a) Sequence of pulses used in the simulation, (b) Population inversion and cavity occupation (dashed blue line) as a function of $gt$. (c) Von Neumann entropy as a function of $gt$. Upper panel shows the evolution of QD qubit over the Poincar\'e sphere, with the color-code indicating the time sequence.}
\label{inv2e}
\end{figure}

Neglecting pure dephasing completely and keeping all parameters as in Fig.\ \ref{inv2d} we obtain the Fig.\ \ref{inv2e}. This two figures have similar general characteristics. The main differences are in range of the population inversion and the initial value of the Von Neumann entropy, which is a result of different initial state as we mention before. In the case with pure dephasing, Fig.\ \ref{inv2d}(b), the population inversion is restricted to smaller absolute values, resulting in self trapping close to the central region of the Poincar\'e sphere, indicating that we have a mixed state. This results in a large value of the Von Neumann entropy, indicating that the initial state prepared (Fig.\ \ref{coh1} for $gt=100$) is not pure. Here it is important to mention that pure dephasing in solid state systems can not be neglected, being always present due to the nature of the system and the presence of phonons, in this way, the preparation of the initial state in solid state system will always face this problem. To minimize its effect, the exciton-cavity detuning can be increased. In systems where pure dephasing can be safely neglected, the result presented in Fig.\ \ref{inv2d} might be useful.

\section{Summary}
\label{sec:conclusion}

In this work, we discuss the dynamics of an exciton in a quantum dot which interacts with a coherent state, supported by a cavity, under the action of a continuous laser, for controlling the cavity losses, and external Gaussian laser pulses, to control exciton-cavity dynamics. Our study is a step towards the implementation of an \textit{on-chip} cavity quantum eletrodynamics. We define a qubit using two levels on the quantum dot so $\ket{0}$ is the vacuum state (no exciton) and $\ket{1}$ is the exciton state (QD qubit). We use the population inversion, $Z(t)$, a mixed vector definition of quantum dot qubit on Poincar\'e sphere to study the dynamics of the qubit and the Von Neumann entropy, to analyze its entanglement degree with the cavity mode (closed system) as well as its purity (open system).

The treatment without considering losses provides important information: by the average population is used to define an efficient condition of qubit-cavity and pulse-cavity detunings for populating the dressed states $\ket{i,n}$. The dynamics shows self-trapping on population inversion, among with oscillations of the relative phase of the qubit.  By including the pulses, we check the effect of the pulse area on the dynamics showing that the central value of population inversion changes significantly after the pulse. The sequence of pulses can be also used to increase the entanglement degree between the qubit and the cavity in non-resonant ideal (no losses) condition.

We also discuss the effect of losses on our approach. Because the QD qubit lifetime is long enough (and it can be even longer in a non-resonant condition with the cavity), we avoid the effects of spontaneous emission by choosing a short time scale defined by $gt<25$. The QD qubit dynamical control, on-demand, is attained by using a sequence of short pulses. To protect the manipulation against cavity looses, we explore the use of an additional continuous laser, which maintains a steady coherent state inside the cavity. This mechanism already sustains the QD qubit dynamics assuring the success of our proposal. We also analyzed the effects of pure dephasing of the QD qubit in the dynamics, showing that it produces no effects in the time evolution during our short pulse sequence, but it might affect the preparation of the initial state.

As future works, we intend to continued studying the entanglement between the QD qubit and cavity by quantifying the existence of quantum correlations when considering losses. The goal is to explore the potential of this system as an entanglement resource. A second issue is to engage a study about production quantum light but focusing on other than single photons.

\section{Acknowledgments}
We would thank the referees for the helpful critics and questions. Their careful and detailed review is behind the enrichment of our work. We acknowledge financial support of CAPES, CNPq, FAPEMIG, DISSE and INCT.

\end{document}